\documentclass[preprint,authoryear,times]{elsarticle}
\usepackage{latexsym}
\usepackage{makeidx}
\usepackage{cite}
\usepackage{amsmath}
\usepackage{color}
\usepackage{upgreek}
\usepackage{morefloats}
\usepackage{setspace}
\usepackage{graphicx}

\begin{document}

\title{Probing Saturn's tropospheric cloud with Cassini/VIMS}
\author[aopp]{J.~K. Barstow\corref{cor1}}
\ead{jo.barstow@physics.ox.ac.uk}
\author[aopp]{P.~G.~J. Irwin}
\author[aopp,leics]{L.~N. Fletcher}
\author[aopp]{R.~S. Giles}
\author[aopp]{C. Merlet}

\cortext[cor1]{Corresponding Author}
\address[aopp]{Atmospheric, Oceanic and Planetary Physics, Clarendon Laboratory, University of Oxford, Parks Road, Oxford, UK}
\address[leics]{Department of Physics and Astronomy, University of Leicester, University Road, Leicester, UK}

\date{Accepted January 2016} 
\copyright {2016. This manuscript version is made available under the CC-BY-NC-ND 4.0 license http://creativecommons.org/licenses/by-nc-nd/4.0/}

\doublespacing

\begin{abstract}
In its decade of operation the Cassini mission has allowed us to look deep into Saturn's atmosphere and investigate the processes occurring below its enshrouding haze. We use Visual and Infrared Mapping Spectrometer (VIMS) 4.6---5.2 $\upmu$m data from early in the mission to investigate the location and properties of Saturn's cloud structure between 0.6 and 5 bars. We average nightside spectra from 2006 over latitude circles and model the spectral limb darkening using the NEMESIS radiative transfer and retrieval tool. We present our best-fit deep cloud model for latitudes $-40^{\circ}<\lambda<50^{\circ}$, along with retrieved abundances for NH$_3$, PH$_3$ and AsH$_3$. We find an increase in NH$_3$ abundance at the equator, a cloud base at $\sim$2.3 bar and no evidence for cloud particles with strong absorption features in the 4.6---5.2 $\upmu$m wavelength range, all of which are consistent with previous work. Non-scattering cloud models assuming a composition of either NH$_3$ or NH$_4$SH, with a scattering haze overlying, fit limb darkening curves and spectra at all latitudes well; the retrieved optical depth for the tropospheric haze is decreased in the northern (winter) hemisphere, implying that the haze has a photochemical origin. Our ability to test this hypothesis by examining spectra at different seasons is restricted by the varying geometry of VIMS observations over the life of the mission, and the appearance of the Saturn storm towards the end of 2010. 
\end{abstract}

\maketitle
\section{Introduction}
It has long been known that clouds are present on the giant planets in our solar system, but attempts to predict their location and composition using microphysical models have so far been relatively unsuccessful (e.g. \citealt{atreya05}, describing the cloud patterns found on Jupiter by the Galileo spacecraft). Clouds are intimately linked with planetary dynamics and chemistry, so understanding their formation and behaviour is a key part of studying any planetary atmosphere. 

The arrival of the Cassini mission at Saturn provided an unprecedented opportunity to study its atmosphere. In the subsequent decade, Saturn's stratospheric composition has been monitored during the changing seasons \citep{fletcher10,sinclair13,fletcher15}; a spectacular hexagonal vortex has been observed at the north pole \citep{fletcher08,baines09}; and the development of a dramatic, large scale storm has been traced over a period of several months \citep{fletcher11b,fischer11,sanchez11,fletcher12,hesman12,sromovsky13,sayanagi13,achterberg14}. Cassini's suite of instruments includes the Visual and Infrared Mapping Spectrometer (VIMS), which provides wavelength coverage between 0.3 and 5.1 $\upmu$m at a spectral resolution of $\sim$16 nm. Absorption bands due to methane, ammonia, phosphine and other trace gases are present in this wavelength range; we can observe the reflected sunlight signature from the dayside at shorter wavelengths, and on the nightside the thermal emission from the planet begins to emerge at around 4.6 $\upmu$m. This broad wavelength coverage provides sensitivity over a large altitude range, making this instrument extremely useful for atmospheric sounding; also, due to the typical size of particles (\citealt{roman13} finds tropospheric haze particles have radii of approximately 2 $\upmu$m), VIMS is highly sensitive beyond 4.6 $\upmu$m to the spectral effect of clouds and haze in the lower atmosphere (between 1 and 8 bars).

In this work, we use VIMS 4.6---5.2 $\upmu$m thermal emission spectra from the nightside of Saturn to investigate the tropospheric cloud and haze. Stratospheric and tropospheric haze properties can be explored using reflected light from the dayside (e.g. \citealt{karkoschka05,sromovsky13,roman13}) but sunlight does not penetrate far enough into Saturn's atmosphere to easily probe cloud much beyond the 1-bar pressure level. On the other hand, thermal emission from the deep atmosphere is absorbed and scattered by clouds in this altitude region \citep{baines06,choi09}, as discussed by \citet{fletcher11}, who presented the first detailed exploration of thermal emission from Saturn using VIMS. They investigated the sensitivity of the spectrum to properties of the tropospheric cloud and haze, as well as determining the latitudinal dependence of PH$_3$, NH$_3$ and AsH$_3$ gas abundances, but found considerable solution degeneracy. We build on this previous work by using spectroscopic limb darkening within latitude circles to provide further constraint on the properties of the cloud and haze. Ground-based observations by \citet{yanamandra01} showed strong latitudinal variation in 5.2 $\upmu$m brightness, attributed to variation in cloud properties; we aim to gain a broad, global picture of Saturn's tropospheric aerosol properties as a function of latitude.

Uncertainty as to the composition and size, and therefore scattering properties, of the Kronian clouds is a major contributor to solution degeneracy in the \citet{fletcher11} study. \citet{atreya05b} use an equilibrium cloud model for Saturn to predict the presence of NH$_3$ ice and solid NH$_4$SH clouds in the troposphere. The NH$_3$ ice cloud is estimated to form a little above the 2 bar level, with the deeper NH$_4$SH cloud forming at around 5 bars. Below this level we may also expect water ice clouds to form, but it is unlikely that these will persist to high enough altitudes for the VIMS measurements to be sensitive to them (at these wavelengths, VIMS is mostly sensitive to pressures between 1 and 8 bar, \citealt{fletcher11}). 

Previous observational work on Saturn's cloud (e.g. \citealt{karkoschka05,fletcher11,sromovsky13,roman13}) has indicated the presence of both stratospheric and tropospheric hazes, with the tropospheric haze located in the region directly above where the NH$_3$ cloud is predicted to form. However, infrared observations sensitive to the deeper troposphere have provided no evidence for the two distinct tropospheric cloud decks (NH$_3$ and NH$_4$SH) above the 10 bar level predicted by \citet{atreya05b} (see \citealt{sromovsky13}). Instead, a single cloud deck beneath the tropospheric haze, in the 1---5 bar range, is preferred, located in between the predicted base pressures for NH$_3$ and NH$_4$SH. This may indicate that the deep cloud is in fact a mixture of these two components, or is composed of either NH$_3$ or NH$_4$SH but also contains impurities. Based on the \citet{atreya05b} model predictions, we consider NH$_3$ and NH$_4$SH compositions for the troposperic cloud and NH$_3$ for the tropospheric haze. Adopting compositions from \textit{ab initio} models in this way reduces the degeneracy of the problem and allows a more informative exploration of other cloud parameters such as particle size.

\section{Data and reduction}
\label{cubes}
\begin{figure*}
\centering
\includegraphics[width=0.8\textwidth]{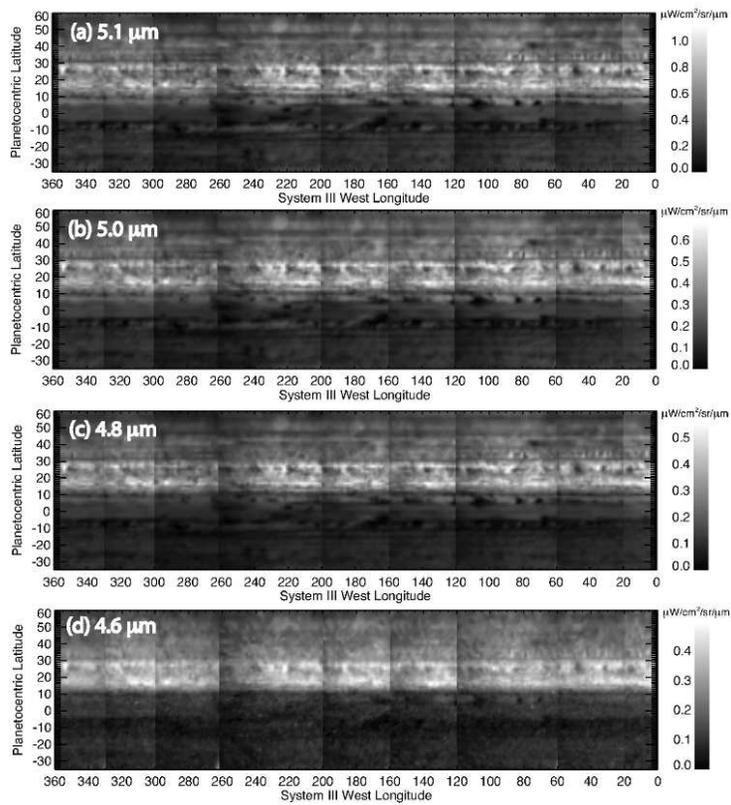}
\caption{Map projected data cubes as used by \citet{fletcher11}, which we also use in this work. A clear hemispherical asymmetry in the 5 $\upmu$m flux is apparent, with the northern latitudes appearing to be much brighter. Reprinted from Icarus, 214, Fletcher, L. N. et al., Saturn's tropospheric composition and clouds from Cassini/VIMS 4.6-5.1 {$\upmu$}m nightside spectroscopy, 510---533, Copyright (2011), with permission from Elsevier\label{leigh_cubes}}
\end{figure*}

We use nightside VIMS cubes from April 2006 (late northern winter/southern summer; Table~\ref{cubes1}) to investigate the cloud limb darkening properties. We choose cubes from this year as Cassini's fairly equatorial orbit at that time allows us to investigate from the equator up to the mid-latitudes of both hemispheres. It also facilitates comparison with \citet{fletcher11}, who used the same cubes. These are overlapping observations taken in a single session while Saturn rotated underneath, such that all longitudes were observed. They are shown in Figure~\ref{leigh_cubes}, in which it can be seen that the northern latitudes are much brighter than the south at 5 $\upmu$m.  

Similar coverage was obtained during 2007, and in Section~\ref{temporal} we compare the limb darkening for the two years. We attempted to investigate further into the mission to see if these trends began to change as Saturn moved towards vernal equinox and into northern summer. However, our ability to do this was restricted by the unavailability of similar data products. Between 2008 and 2010, we could not locate a sequence of VIMS images covering a wide simultaneous range of latitudes and emission angles, which is required for a study of this kind. This was due to the spacecraft moving to an inclined orbit. Towards the end of 2010 the large Saturn storm emerged, causing great disruption to the atmosphere with effects that persisted for several Earth years, preventing any further study.

\begin{table}
\centering
\begin{tabular}[c]{|c|c|c|}
\hline
Observation&Date&Integration Time (s)\\
\hline
CM1524383985 & 2006-04-22 & 480 \\
CM1524388848 & 2006-04-22 & 480 \\
CM1524393612 & 2006-04-22 & 480 \\
CM1524400806 & 2006-04-22 & 480 \\
CM1524403247 & 2006-04-22 & 480 \\
CM1524408018 & 2006-04-22 & 480 \\
CM1524412815 & 2006-04-22 & 480 \\
CM1524417617 & 2006-04-22 & 480 \\
\hline
\end{tabular}
\caption{List of 2006 data cubes used in the current research. \label{cubes1}}
\end{table}

\begin{figure*}
\centering
\includegraphics[width=0.8\textwidth]{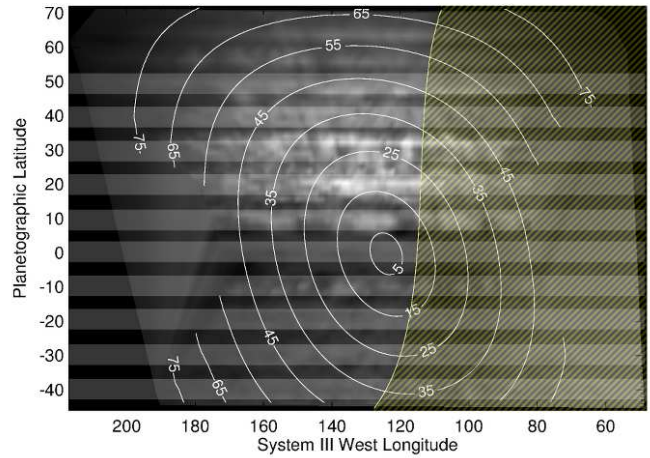}
\caption{Map projected data cube CM1524383985, shown to illustrate our data selection procedure. White contours show emission angles. Hatching indicates region rejected due to sunlight contamination (solar incidence angle less than 105$^{\circ}$). Pale shaded stripes show the latitude circles used. \label{data_north}}
\end{figure*}

We investigate the latitudinal dependence of the cloud properties by exploiting the change in emission angle along a latitude circle as viewed by VIMS. To first order, we do not expect significant longitudinal variability at these pressure levels on Saturn (see \citealt{yanamandra01} for details of 5 $\upmu$m variability observed from the ground), so much of the broad zonal variation evident in Figures~\ref{leigh_cubes} and~\ref{data_north} is due to limb darkening. To average out small-scale longitudinal variation we take all 8 data cubes together and bin the spectra in latitude and emission angle. Any data points with a solar incidence angle of less than 105$^{\circ}$ are rejected, to avoid contamination from reflected sunlight. We extract latitude circles from 40$^{\circ}$S to 50$^{\circ}$N, taking all pixels at intervals of 10$^{\circ~}$ in latitude with a spread of $\pm$3$^{\circ}$. By binning over a broad latitude band we hope to average over any small-scale variation, although it means that we do not resolve the fine details of the latitude variation as reported by \citet{fletcher11}.

\begin{figure*}
\centering
\includegraphics[width=0.8\textwidth]{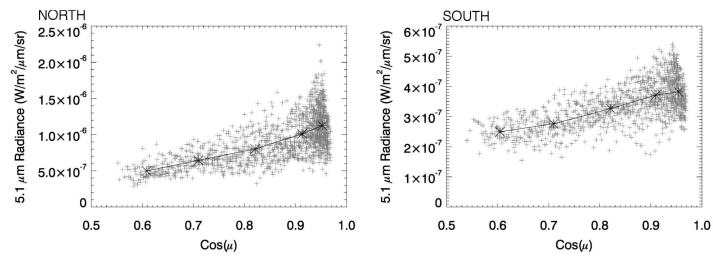}
\caption{5.1 $\upmu$m radiance for all pixels within a single latitude circle at 20$^{\circ}$ north and south. Whilst pixel-level variations are apparent, the shape of the limb darkening relation is well represented by the average. This encompasses both latitudinal and longitudinal variation.\label{limb_nobin}}
\end{figure*}

We also average spectra in the longitudinal direction every 10$^{\circ}$ in emission angle, to ensure that any local features are smoothed out. This results in an emission angle range between 5$^{\circ}$ and 45$^{\circ}$ at the equator, and 35$^{\circ}$ and 85$^{\circ}$ at the highest latitudes. The average limb darkening at 5.1 $\upmu$m for latitude circles and 20$^{\circ}$ north and south are shown in Figure~\ref{limb_nobin}. The small scale variation is apparent, but it is also clear that the average captures the basic limb darkening trend well. The variation in emission angle range is due to the equatorial location of the spacecraft during these observations, leading to generally higher emission angles further from the equator. At low and equatorial latitudes, the images do not extend to the limb of the planet, truncating the limb darkening range. An example of the manipulation of a single cube is shown in Figure~\ref{data_north}. 

The cubes were downloaded from the NASA PDS archive and calibrated using the standard ISIS pipeline \citep{fletcher11}. The cubes are projected onto a System III planetographic latitude/longitude grid. Radiometric errors are conservatively estimated to be 12\% of the average flux between 4.7 and 5.1 $\upmu$m; the 12\% value is based on the error estimates used in \citet{fletcher11}, but we assume a constant error of 12\% of the average 4.7---5.1 $\upmu$m flux across all wavelengths and all emission angles,  which favours the spectral regions where the signal is largest. 

\begin{table}
\centering
\begin{tabular}[c]{|c|c|}
\hline
Latitude ($^{\circ}$)& Emission angle range ($^{\circ}$)\\
\hline
-40 & $\sim$35---85\\
-30 & $\sim$25---65\\
-20 & $\sim$15---55\\
-10 & $\sim$5---55\\
0  & $\sim$5---45\\
10 & $\sim$5---55\\
20 & $\sim$15---55\\
30 & $\sim$25---65\\
40 & $\sim$35---85\\
50 & $\sim$45---85\\
\hline
\end{tabular}
\caption{Emission angle ranges for each latitude circle. The ranges refer to the central angle of the highest or lowest 10$^{\circ}$ range included for each latitude circle. The angles increase towards higher latitudes because the sub-spacecraft point lies close to the equator for all observations. \label{cloudmods}}
\end{table}


\section{Model atmosphere and retrieval}
\begin{figure}
\centering
\includegraphics[width=0.75\textwidth]{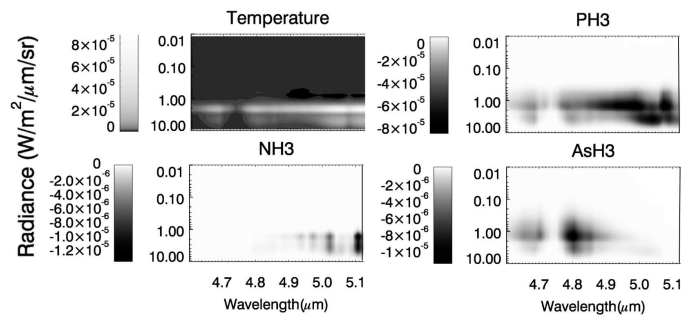}
\caption{Jacobians (functional derivatives) for temperature, PH$_3$, NH$_3$ and AsH$_3$ for a typical cloudy model atmosphere as used in this work. A compact tropospheric cloud is located at 2.3 bars and an extended haze between 0.1 and 0.6 bar. Jacobians show sensitivity to changes in temperature (per K) and PH$_3$, NH$_3$ and AsH$_3$ abundances  (per log volume mixing ratio) at different altitudes. The VIMS instrument is mostly sensitive to pressures between 1 and 8 bar. The effect of the cloud can be clearly seen in the increased sensitivity above the 2.3-bar level in the temperature and PH$_3$ Jacobians, whereas the haze is too high up to have a similar effect. \label{jacobians}}
\end{figure}

The Saturn model atmosphere is based on the work of \citet{fletcher11}. Line data sources are as in this previous paper and \citet{giles15}. After \citet{fletcher11}, we use latitudinally varying temperature profiles based on retrievals from the Composite Infrared Spectrometer (CIRS) instrument averaged over the 2004---2008 period of observations, extrapolated to an adiabat in the deep atmosphere \citep{fletcher07,fletcher10}. CIRS (FP3/FP4) operates in the 7---16 $\upmu$m wavelength range, making it highly sensitive to Saturn's thermal emission from 600---100 mb and 10---1 mb and therefore an ideal probe of temperature in the upper troposphere and middle stratosphere. The temperature at higher pressures, closer to the regions in which VIMS is sensitive (see Figure~\ref{jacobians}), is not probed by the CIRS instrument and cannot be independently constrained using VIMS data; however, we expect the temperature to be more stable in the deeper regions of the atmosphere. We do not, therefore, expect small-scale variability to affect our results, especially as we average over spectra from eight different data cubes.

\subsection{Retrieval Algorithm}
We use the NEMESIS radiative transfer and retrieval algorithm \citep{irwin08} to simultaneously retrieve several atmospheric properties from the VIMS spectra. After \citet{fletcher11}, we only vary the model parameters to which the 4.6---5.1 $\upmu$m spectral region is most sensitive, keeping everything else fixed. The main absorbers in this spectral region are the cloud and haze, as discussed in further detail in Section~\ref{cloudsection}. Regarding molecular absorbers, \citet{fletcher11} found that, whilst NH$_3$, PH$_3$, AsH$_3$, GeH$_4$, CH$_4$ and CH$_3$D have absorption in this region, variation in the abundances of GeH$_4$, CH$_4$ and CH$_3$D had an insignificant effect on spectra at the resolution of VIMS. The abundances of these gases were therefore fixed to the values used in \citet{fletcher11}, leaving only three variable gases. The sensitivity of the spectrum to absorption by PH$_3$, NH$_3$ and AsH$_3$ is indicated in Figure~\ref{jacobians}, with PH$_3$ having the broadest effect across the spectrum. NH$_3$ absorbs at wavelengths longer than 4.8 $\upmu$m and AsH$_3$ at shorter wavelengths. 

\begin{figure*}
\centering
\includegraphics[width=0.85\textwidth]{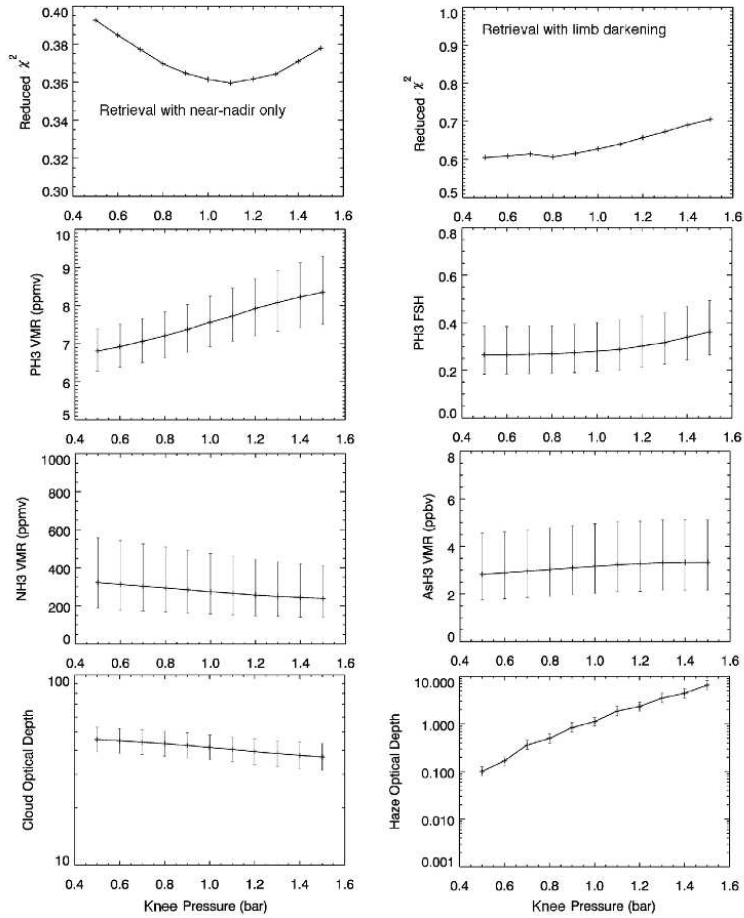}
\caption{Retrieved atmospheric properties and fitting accuracies for limb darkening relations at 20$^{\circ}$S when different PH$_3$ knee pressures are used. Pressures around 1.1 bar produce the best fit if only the nearest-nadir spectrum is considered (top left-hand $\chi^2$ plot), but the full limb darkening series slightly favours a lower pressure closer to the CIRS value (top right-hand $\chi^2$ plot). However, we find that most retrieved values depend very little on the chosen knee pressure. \label{ph3knee20S}}
\end{figure*}

\citet{fletcher11} found that the CIRS-derived PH$_3$ profile did not provide a good fit to the VIMS data. The CIRS instrument focal planes 3 and 4, previously used to constrain PH$_3$ abundance \citep{fletcher09}, are sensitive to absorption by PH$_3$ at lower pressures (300---800 mbar) than the VIMS measurements. As \citet{fletcher11}, in this work the PH$_3$ profile is modelled with a constant volume mixing ratio up to a given pressure level (the `knee pressure'), above which the abundance drops off as a function of altitude. CIRS measurements indicated that PH$_3$ would be well-mixed up to 0.55 bar, but \citet{fletcher11} found, when analysing the VIMS observations, that the knee pressure instead occurred at 1.3 bar. We test the effect on the retrieval of varying this knee pressure. Though deeper pressures for the knee between 1.1 and 1.5 bar do indeed produce a better fit for the single nearest-to-nadir spectra, lower pressures produce a slightly better fit for limb darkening profiles, with a lower reduced $\chi$-squared\footnote{The $\chi^2$ goodness-of-fit parameter is calculated using the relation $\chi^2 = (\Sigma (y_{measured}-y_{modelled})^2/\sigma^2)$; the reduced $\chi^2$ is this sum divided by the number of degrees of freedom.} (Figure~\ref{ph3knee20S}). The $\chi^2$ for the limb darkening is calculated by treating all spectra in the limb darkening sequence as a single dataset. However, the spectral shape is reproduced slightly less well at these lower pressures. Changing the knee pressure does result in some changes in retrieved values: the tropospheric PH$_3$ abundance decreases for lower knee pressures, to compensate for the fact that the tropospheric abundance is fixed up to a higher altitude; the cloud optical depth decreases slightly whilst the haze optical depth increases, and a similar trade-off is seen between NH$_3$ and AsH$_3$, with NH$_3$ decreasing and AsH$_3$ increasing. However, all of these effects are small and in the majority of cases do not exceed the retrieval error. We therefore fix the knee pressure at 1.3 bar to facilitate simple comparison with the results of \citet{fletcher11}, and retrieve the deep PH$_3$ abundance and a fractional scale height above the knee. 

\textit{A priori} abundances for PH$_3$, NH$_3$ and AsH$_3$ are the best-fit values from \citet{fletcher11}, and the other model atmosphere parameters are the same, with the exception of the cloud properties used which are explained further in Section~\ref{cloudsection}. We retrieve scaled specific densities for the cloud species included in the model.

\subsection{Cloud models}
\label{cloudsection}
The key differences between this work and that of \citet{fletcher11} are in the treatment of clouds, and the use of limb darkening relations to place further constraints on their properties. \citet{fletcher11} do not consider limb darkening and the majority of their conclusions are based on nadir geometry spectra. Because of the different paths through the atmosphere for spectra at different emission angles, using limb darkening relations provides further information. For example, if particles are strong scatterers then the limb darkening effect is less pronounced than for mostly-absorbent particles, as more light propagates through the atmosphere at low emission angles. \citet{roos06,giles15} demonstrate that scattering is important in the Jovian atmosphere at 5 $\upmu$m; we test the requirement for a scattering tropospheric cloud layer on Saturn by comparing scattering and non-scattering retrievals. The retrieved atmospheric properties depend on both the cloud vertical structure and scattering properties, with limb darkening providing a tighter constraint on both than single observations taken close to the nadir. 

\citet{fletcher11} consider three compositions for the tropospheric cloud: a non-scattering grey cloud, an NH$_3$ ice cloud with refractive indices taken from \citet{martonchik84}, and an NH$_4$SH cloud with refractive indices taken from \citet{howett07}. Extended and compact, scattering- and non-scattering variants of these models were considered, with or without the presence of an overlying grey cloud at the condensation pressure expected for the NH$_3$ cloud (around 1.5 bar).

 We further explore the effect of different cloud scattering properties on VIMS spectra by considering the effects on limb darkening relationships. To this end, we consider four different cloud model scenarios (Table~\ref{cloudmodels}). These consist of a compact deep cloud made of either NH$_3$ or NH$_4$SH extending over a single model layer, as \citet{fletcher11} found that a compact tropospheric cloud produced a better fit than an extended cloud. In two models we also include a `tropospheric haze' layer as indicated by the Cassini ISS analysis of \citet{roman13}; we choose to consider a haze layer rather than the 2-cloud model adopted by \citet{fletcher11} as there is evidence for the presence of a tropospheric haze layer from reflected-light observations of Saturn, located higher up than a compact layer at 1.5 bar.  This haze layer is extended between 0.1 and 0.6 bar, with an effective particle radius of 2 $\upmu$m, as suggested by \citet{roman13}. This is also consistent with the results of \citet{karkoschka05}. \citet{roman13} do not suggest a composition for this haze, but use a real refractive index of $\sim$1.43 at visible wavelengths, compatible with \citet{martonchik84} values for NH$_3$ ice. Likewise, we use NH$_3$ refractive index properties for this haze, since we know that NH$_3$ is abundant on Saturn and it is expected to condense, albeit at slightly higher pressures than the tropospheric haze layer.

\begin{figure*}
\centering
\includegraphics[width=1.0\textwidth]{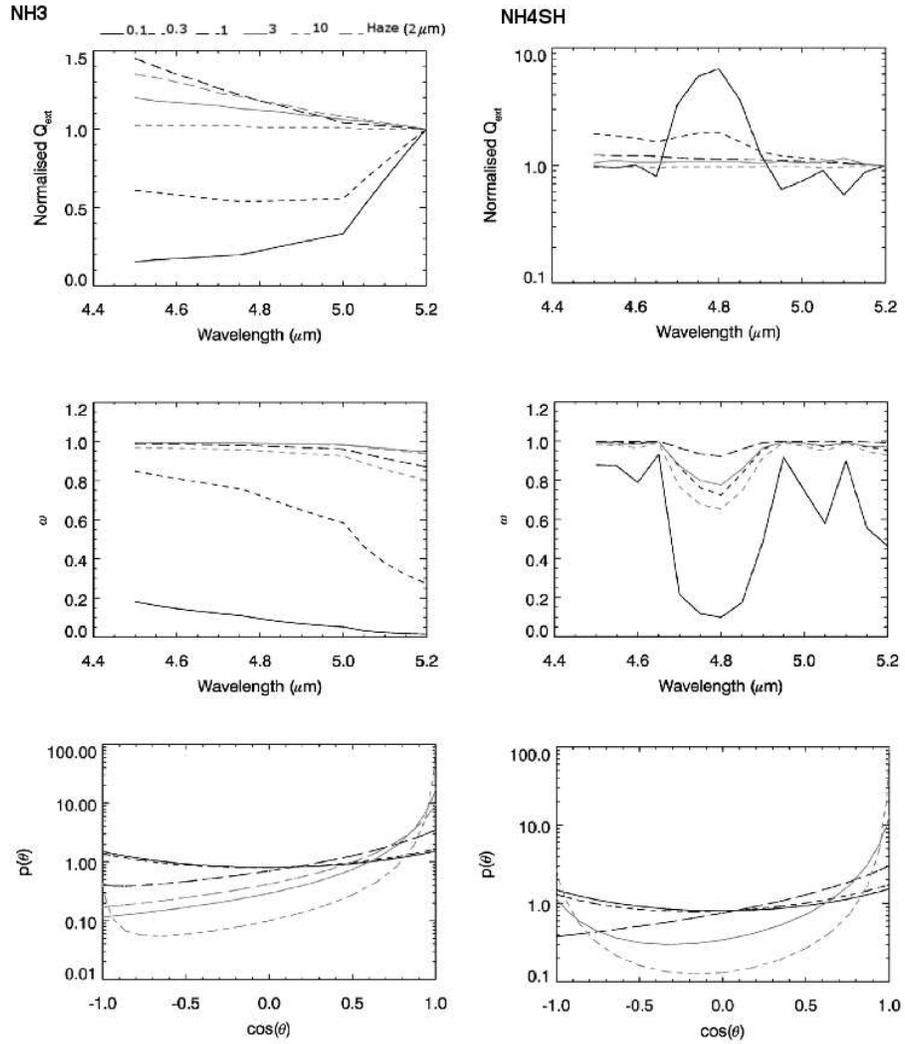}
\caption{Extinction cross section (normalised to unity at 5.2 $\upmu$m), single-scattering albedo, and phase function at 5.1 $\upmu$m for NH$_3$ and NH$_4$SH clouds of different particle size distributions, where the effective radius of is shown in the key to the top left.\label{ext}}
\end{figure*}

The extinction cross section and single-scattering albedo are relatively uniform for 2 $\upmu$m NH$_3$ particles over the 5$\upmu$m wavelength range, as can be seen in Figure~\ref{ext}. At 5 $\upmu$m, larger NH$_3$ particle sizes have more spectrally uniform properties compared with smaller particles. Together with the evidence from \citet{roman13}, this spectral invariance is why we choose to use 2 $\upmu$m haze particles; given that the composition of the haze is unknown, we do not want to introduce spurious absorption features into the spectrum when haze is included.

We choose to use properties for NH$_3$ and NH$_4$SH particles rather than an arbitrary set of cloud properties. Introducing clouds adds a number of free parameters to the model, and it is clear from the results of \citet{fletcher11} that the problem is very degenerate. This is therefore the simplest scenario, although more complicated ones exist, and these could be explored by allowing the refractive indices of the cloud constituent to be free parameters. However, this approach would extend the parameter space for this work to an unfeasibly wide range for a single study, and would be unlikely to enable us to make any more meaningful statements about the cloud properties. NH$_3$ ice and NH$_4$SH are the two species that are predicted to condense at pressures to which the VIMS 5-$\upmu$m measurements are sensitive, hence our choice. In the case where the data can be represented by a cloud made of either species, the retrieved cloud base pressure may serve as an indicator of composition should it occur at a predicted cloud base pressure for either NH$_3$ or NH$_4$SH.

\begin{table}
\centering
\begin{tabular}[c]{|c|c|c|}
\hline
& NH$_3$ compact cloud & NH$_4$SH compact cloud\\
\hline
No haze & A & B \\
NH$_3$ tropospheric haze & C & D \\
\hline
\end{tabular}
\caption{Four different cloud models used in this work. The fifth model (E) is a simple grey cloud model.\label{cloudmodels}}
\end{table}

For models A---D, we test five different particle sizes for the cloud: 0.1, 0.3, 1, 3, and 10 $\upmu$m. The spectral properties for each of these, and for the 2 $\upmu$m particles we use for the haze model, are shown in Figure~\ref{ext}. The significant changes in scattering properties over this range should allow some constraint to be placed on  particle size. For each size, a gamma distribution with a variance of 0.05 is used; this provides a relatively tight size distribution, but is broad enough to wash out smaller scale spectral features that arise when monodisperse particle size distributions are considered, making the extinction cross section and single-scattering albedo curves smooth. \citet{fletcher11} used the same size distribution, but only considered 1 $\upmu$m particles for the tropospheric cloud. We compute the scattering cross-sections, single-scattering albedos and phase functions using Mie theory, assuming spherical particles. The calculated phase functions are then approximated using a two-term Henyey-Greenstein function.

For each of the models we also vary the base pressure between 0.7 and 5 bar. For each case, we perform a separate retrieval of the gas variables, cloud optical depth (and tropospheric haze optical depth if present) and compare the goodness of the resulting fit using the reduced $\chi^2$ statistic. To select the best-fit cloud size and base pressure for each model we perform a full retrieval for the lowest emission angle spectrum within each latitude circle, and then use the retrieved parameters to forward model the limb darkening, as a full retrieval using all spectra is very computationally intensive. Once the best-fit cloud parameters are selected, full retrievals over all limb darkening spectra are performed for that case only. Retrieved quantities and parameterizations are listed in Table ~\ref{rets}.

\begin{table}
\centering
\begin{tabular}[c]{|c|c|}
\hline
Retrieved quantity & Parameterization\\
\hline
Cloud specific density & log multiple of profile (converted to optical depth)\\
Haze specific density & log multiple of profile (converted to optical depth)\\
PH$_3$ VMR & Deep abundance, fractional scale height above knee pressure\\
NH$_3$ VMR & log multiple of profile\\
AsH$_3$ VMR & log multiple of profile\\
\hline
\end{tabular}
\caption{Retrieved quantities and methods of parameterization. Cloud particle size and base pressure are varied manually between retrievals.\label{rets}}
\end{table}

\section{Results}
\label{results_section}
\begin{figure}
\centering
\includegraphics[width=0.75\textwidth]{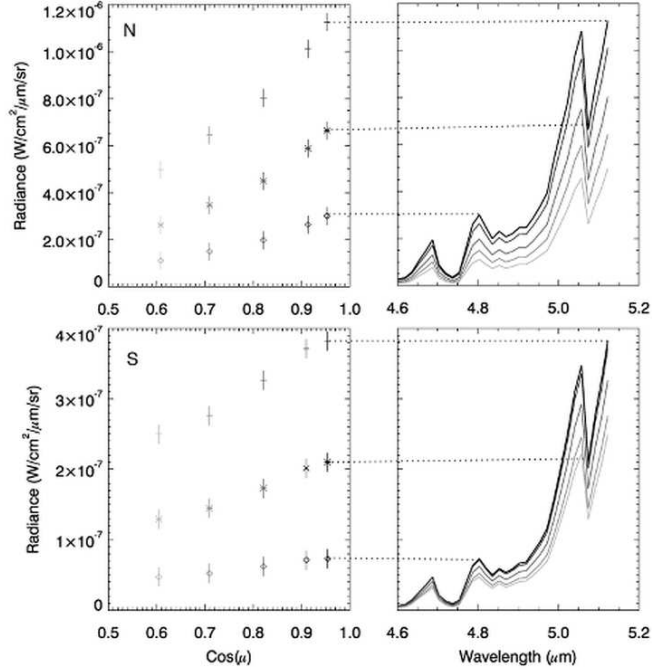}
\caption{Spectra as a function of emission angle and corresponding limb darkening curves for the 20$^{\circ}$N and 20$^{\circ}$S data. The shades of grey correspond to spectra at different emission angles. The symbols in the limb darkening curves on the left refer to the wavelengths indicated on the bold spectrum in the right-hand panels.\label{20S_data}}
\end{figure}

\begin{figure}
\centering
\includegraphics[width=0.75\textwidth]{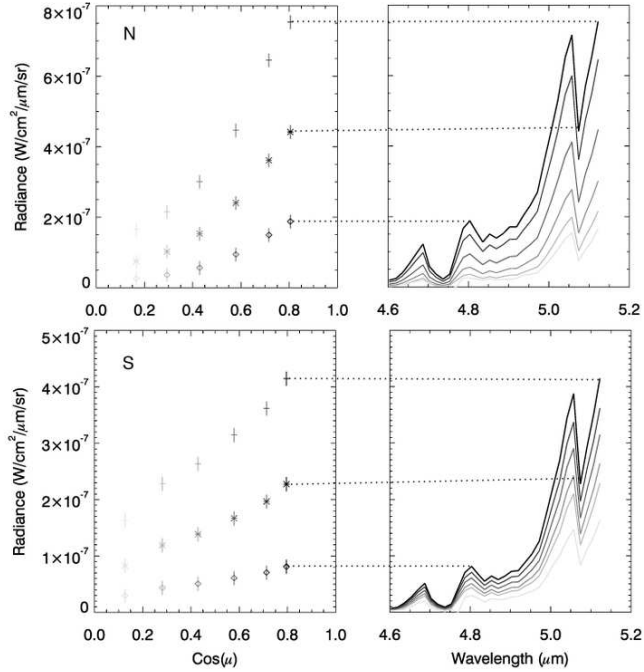}
\caption{As Figure~\ref{20S_data}, but for spectra at 40$^{\circ}$N and 40$^{\circ}$S. As in Figure~\ref{20S_data}, the limb darkening curves in the north are steep compared with the limb darkening curves in the south.\label{40S_data}}
\end{figure}

 We present results from our range of retrieval tests here. It is immediately clear from consideration of the data that the limb darkening relationships differ between the southern and northern hemispheres. \citet{fletcher11} identified hemispheric differences, with the northern hemisphere consistently appearing to be brighter, indicating that these latitudes are comparatively less cloudy, and we also see a stronger limb darkening slope in comparison with the southern hemisphere measurements. The change in brightness immediately suggests that the cloud and/or haze is optically thinner in the north - \citet{baines06} estimate that the 5$\upmu$m optical depth in the north is about 0.7 less than that in the south. This behaviour is seen across the 4.6---5.2 $\upmu$m window. In Figure~\ref{20S_data}, we compare the observations at 20$^{\circ}$N and 20$^{\circ}$S, and this effect is clear. Similar behaviour is also seen at other pairs of latitudes, e.g. 40$^{\circ}$N and 40$^{\circ}$S (Figure~\ref{40S_data}).

\begin{table}
\centering
\begin{tabular}[c]{|c|c|c|}
\hline
& NH$_3$ compact cloud & NH$_4$SH compact cloud\\
\hline
Scattering cloud & A & B \\
Absorbing cloud  & Aiii & Biii \\
Scattering cloud + haze & C & D \\
Absorbing haze, scattering cloud & Ci & Di \\
Absorbing cloud, scattering haze & Cii & Dii \\
Absorbing cloud + haze  & Ciii & Diii \\

\hline
\end{tabular}
\caption{Scattering variants of cloud models A, B, C and D\label{cloudmods2}}
\end{table}

Steeper limb darkening is indicative of cloud or haze that is more efficient at absorbing and less efficient at scattering light. The longer path length at high emission angles has an increased effect if the atmosphere is more absorbing, producing greater attenutation at high emission angles. We therefore test variants of cloud models A, B, C and D to investigate possible causes of increased limb darkening. These variants are listed in Table~\ref{cloudmods2}. To achieve a non-scattering haze/cloud only, we set the single-scattering albedo for haze/cloud at all wavelengths to zero. Where both the cloud and the haze are non-scattering, we simply run the retrieval with scattering turned off. 

The only variant of cloud models C and D that produce a good fit at all latitudes is the case where the deep cloud is assumed to be non-scattering but the haze is not (models Cii and Dii). It is possible to reproduce the different limb darkening relationships with a single model because the retrieved haze optical depth is lower in the northern (winter) hemisphere than the southern hemisphere. Therefore, the combination of cloud and haze in the southern hemisphere is more scattering overall than the northern hemisphere cloud and haze, which produces shallower limb darkening in the southern hemisphere compared with the northern hemisphere. 

The difference in goodness of fit between NH$_4$SH and NH$_3$ compositions for the tropospheric cloud is very small, although for most latitudes a slightly better fit is obtained for NH$_4$SH. This is a result of the fact that for either case particle sizes are favoured for which the extinction cross section and single-scattering albedo variation with wavelength is small -- no strong spectral features of the cloud are visible in the spectrum. To test the effect of spectral features due to the cloud, we test a further model (cloud model E) which is based on models Cii and Dii but has spectrally-invariant properties for the cloud and haze. The cloud becomes a simple grey, non-scattering cloud, and the haze is grey and scattering with a spectrally invariant phase function (based on the phase function for 2 $\upmu$m NH$_3$ particles at 5 $\upmu$m, as shown in Figure~\ref{ext}). 

\begin{figure*}
\centering
\includegraphics[width=0.85\textwidth]{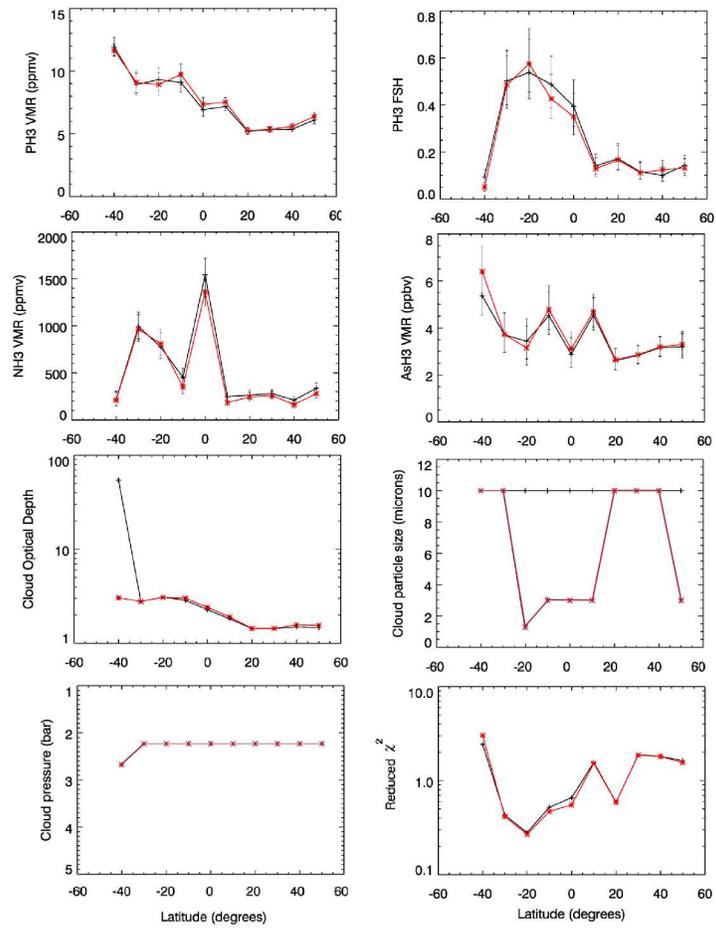}
\caption{Best-fit retrieved values for each latitude circle for models Aiii - black crosses and Biii - red stars. Cloud optical depth is quoted at 5.1 $\upmu$m.\label{results_nohaze}}
\end{figure*}

\begin{figure*}
\centering
\includegraphics[width=0.85\textwidth]{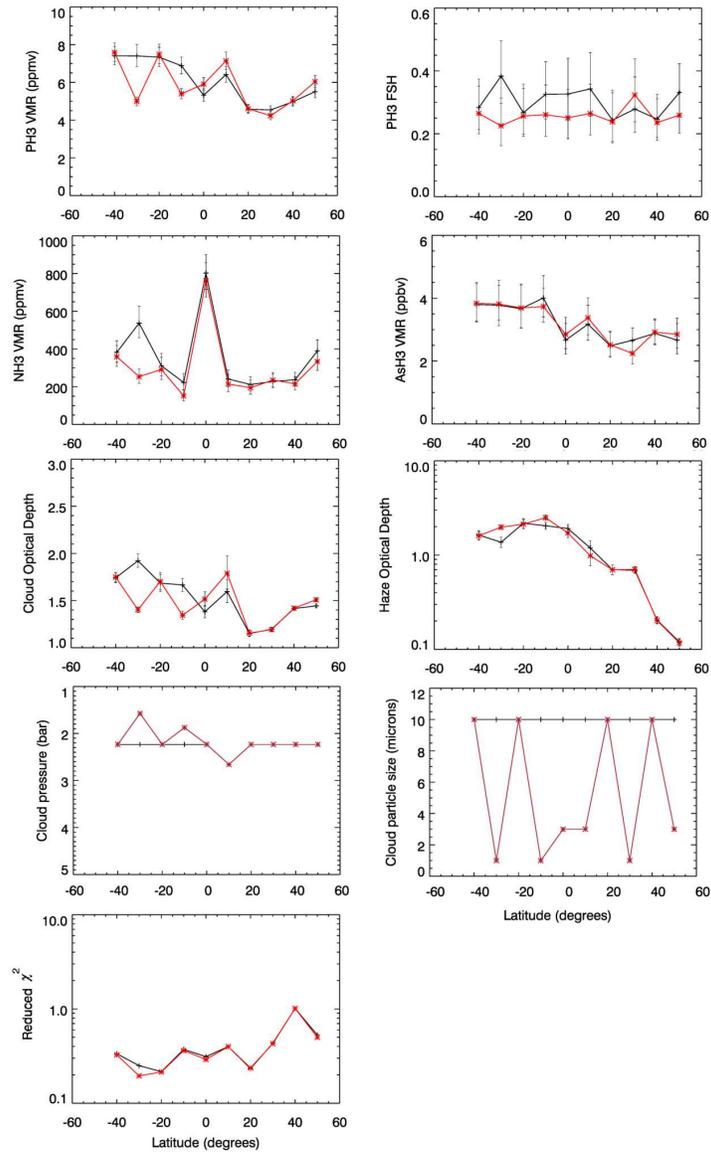}
\caption{Best-fit retrieved values for each latitude circle for models Cii - black crosses and Dii - red stars. The jumps in cloud base pressure and particle size occur because a range of discrete pressures and sizes is tested. Cloud/haze optical depth are quoted at 5.1 $\upmu$m. \label{resultsall}}
\end{figure*}

\begin{figure*}
\centering
\includegraphics[width=0.85\textwidth]{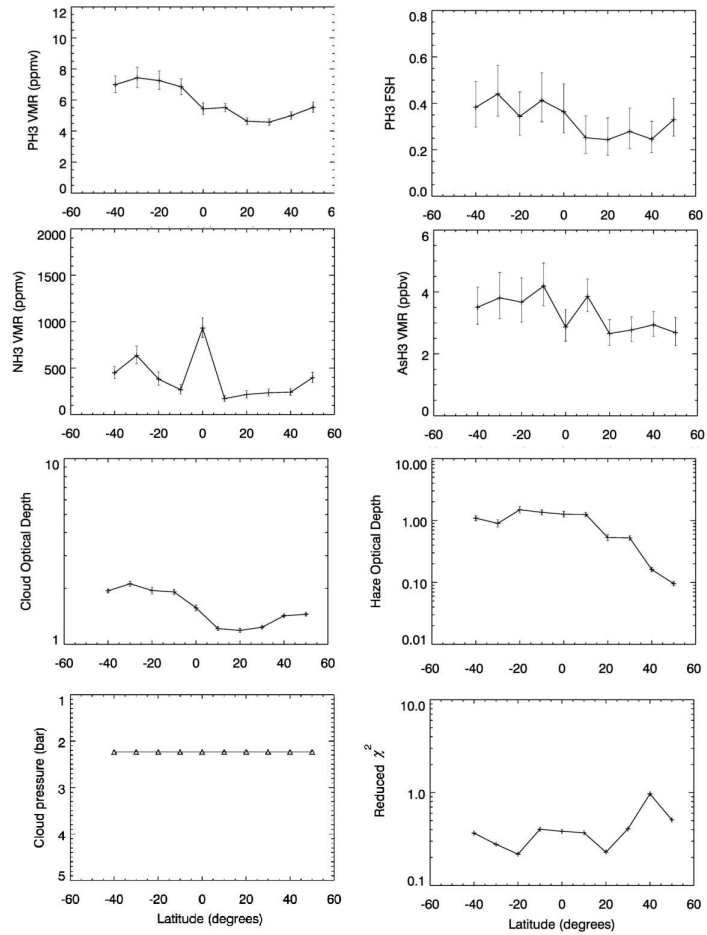}
\caption{Best-fit retrieved values for each latitude circle for model E (triangles). Cloud/haze optical depth are quoted at 5.1 $\upmu$m.\label{resultsgrey}}
\end{figure*}

We find that in the majority of cases models including haze (C and D) provide a slightly better fit to the spectra than models without haze (A and B). The exception is at 50$^{\circ}$N where models A and B provide the best fit, and where the retrieved haze optical depth for models C and D is low anyway. The best-fit model variants for NH$_3$ and NH$_4$SH compositions are shown in Table~\ref{cloudmods3}. The fact that haze models are generally favoured is not an unexpected result as the tropospheric haze has been identified from observations at shorter wavelengths, as in e.g. \citet{roman13}. This fact, coupled with the need for the haze to reproduce the change in limb darkening properties as a function of latitude, suggests that the VIMS thermal emission spectral region is sensitive to both a tropospheric cloud and to some opacity higher in the atmosphere. The retrieved parameters for the best-fit A and B models are presented in Figure~\ref{results_nohaze}, for best-fit C and D models in Figure~\ref{resultsall}, and for best-fit non-scattering grey cloud/scattering grey haze models (E) in Figure~\ref{resultsgrey}.

\begin{figure*}
\centering
\includegraphics[width=0.85\textwidth]{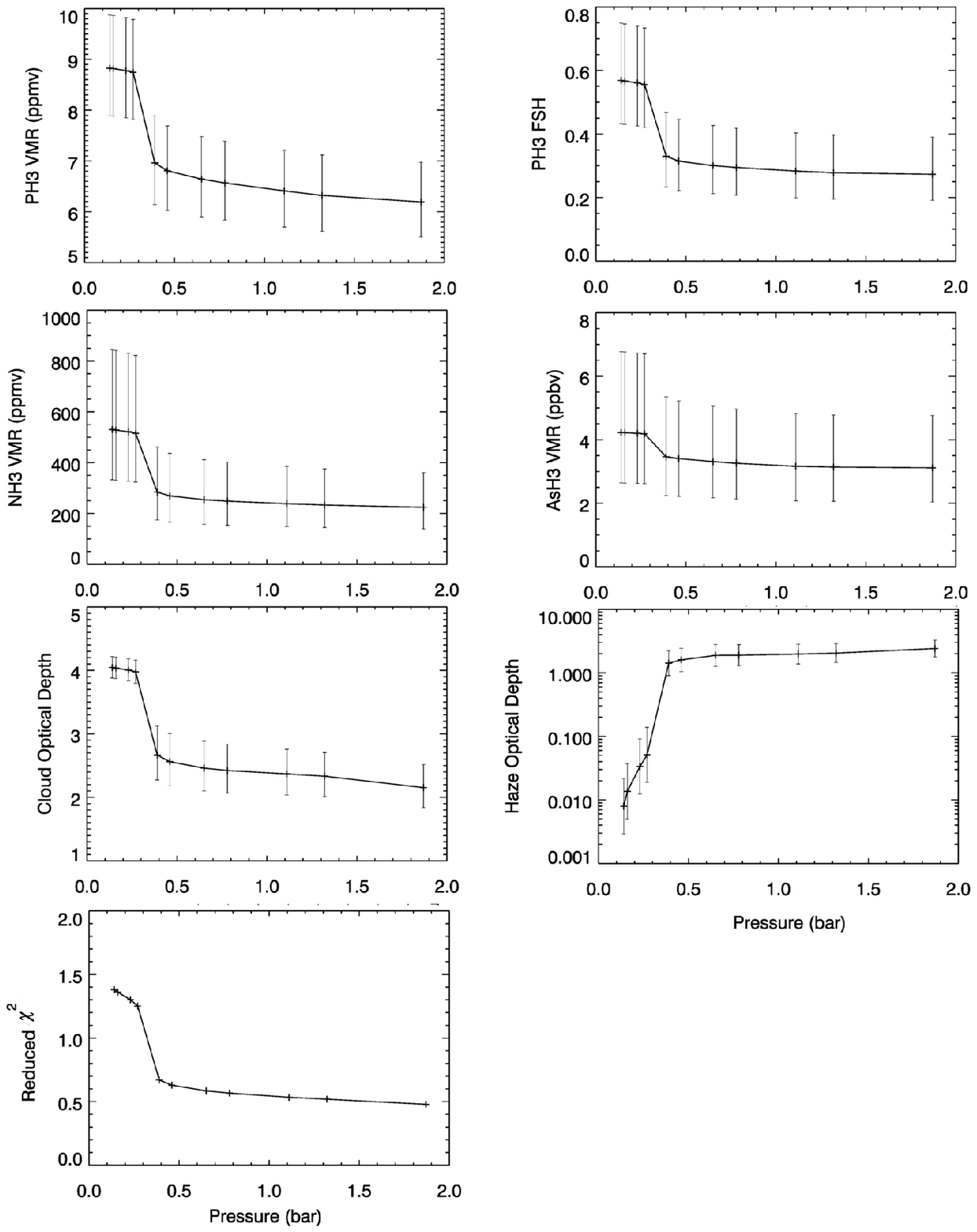}
\caption{Variation in retrieved quantities for different haze base pressures at 20$^{\circ}$S. Base pressures lower than 0.4 bar do not provide as good a fit to the spectrum, indicating that the haze must be extended and not compact (the haze top pressure is 0.1 bar). For base pressures greater than 0.4 bar, the variation in goodness of fit and retrieved quantities is small and within error.\label{hazebase}}
\end{figure*}

For the $\pm$20$^{\circ}$ latitude circles, we test different base pressures for the haze between 1.8 and 0.2 bar (for a tropospheric cloud base pressure of 2.3 bar; Figure~\ref{hazebase}). For all base pressures higher than 0.4 bar, we find there is good fit to the spectra, and the variation in retrieved values for different haze base pressures is within the error bars and therefore insignificant; the top and base pressures are therefore fixed at 0.1 and 0.6 bar respectively, which are values taken from \citet{roman13} and constrained by reflection results. This base pressure is consistent with other literature values, including \citet{stam01}, \citet{munoz04}, and \citet{carlson10}. It should be noted that, as deeper base pressures for the haze are not excluded, there is no requirement for a gap between the tropospheric cloud and haze; however, the findings relating to scattering and the strong variation in haze optical depth with latitude demonstrates that the cloud and haze must be independent of each other. Conversely, there is strong evidence that the haze must be an extended layer rather than a compact layer, as the goodness of fit gets significantly worse as the haze base pressure gets closer to the top pressure.

\begin{table}
\centering
\begin{tabular}[c]{|c|c|c|c|c|c|c|c|c|c|c|}
\hline
& A & $\chi^2_{red}$ & B & $\chi^2_{red}$ & C & $\chi^2_{red}$ & D & $\chi^2_{red}$ & E & $\chi^2_{red}$\\
\hline
50$^{\circ}$ & Aiii & 0.82 & \textbf{Biii} & \textbf{0.72} & Cii & 1.2 & Dii & 0.94 & E & 1.4\\
40$^{\circ}$ & Aiii & 1.6 & Biii & 1.5 & Cii & 1.5 & Dii & 1.4 & \textbf{E} & \textbf{1.4}\\
30$^{\circ}$ & Aiii & 1.6 & Biii & 1.6 & \textbf{Ci} & \textbf{0.38} & Di & 0.48 & E & 1.4 \\
20$^{\circ}$ & Aiii & 0.73 & Biii & 0.72 & \textbf{Ci} & \textbf{0.53} & Dii & 0.63 & E & 0.61\\
10$^{\circ}$ & Aiii & 1.5 & Biii & 1.4 & Ci & 1.2 & \textbf{Dii} & \textbf{1.0} & E & 1.4 \\
0$^{\circ}$ & A & 0.68 & Biii & 0.68 & \textbf{Ci} & \textbf{0.56} & Di & 0.66 & E & 0.79\\
-10$^{\circ}$ & Aiii & 0.75 & Biii & 0.69 & Cii & 0.63 & \textbf{Dii} & \textbf{0.50} & E & 0.67  \\
-20$^{\circ}$ & Aiii & 0.80 & Biii & 0.80 & Cii & 0.29 & \textbf{Dii} & \textbf{0.29} & E & 0.32\\
-30$^{\circ}$ & Aiii & 1.2 & Biii & 1.2 & Cii & 0.76 & \textbf{Dii} & \textbf{0.19} & E & 1.1\\
-40$^{\circ}$ & Aiii & 1.6 & Biii & 1.6 & Cii & 0.39 & \textbf{Dii} & \textbf{0.37} & E & 0.47\\
\hline
\end{tabular}
\caption{Best fit variants of the five cloud models for each latitude; models A/B are NH$_3$/NH$_4$SH models without haze, models C/D are the same with haze. Models i) have scattering cloud and non-scattering haze, models ii) have scattering haze and non-scattering cloud, and models iii) are completely non-scattering.  Model E is a grey model with scattering haze and non-scattering cloud. The best-fitting models for each latitude are highlighted using bold font. Quoted reduced $\chi^2$ values are for fits to the nearest-nadir spectrum with forward modelled limb darkening.\label{cloudmods3}}
\end{table}

It is obvious from Table~\ref{cloudmods3} that the parameter space is highly degenerate; however, there are broad trends that can be immediately extracted. Models including haze are generally favoured. The difference in goodness of fit between the hazy and haze-free models is most clear at southern latitudes, which makes intuitive sense as this is where the retrieved haze optical depth is greatest. 

\begin{figure}
\centering
\includegraphics[width=0.85\textwidth]{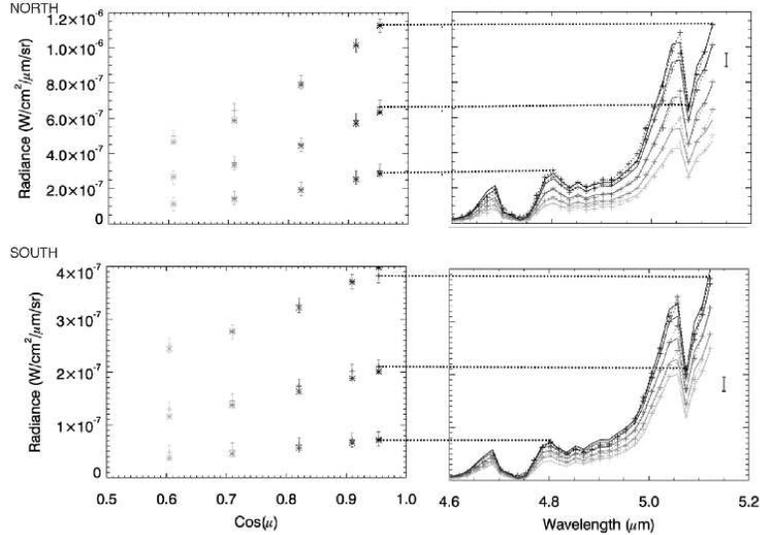}
\caption{Full spectral fits at 20$^{\circ}$N and 20$^{\circ}$S, for full retrievals using the cloud model Dii. Left: limb darkening data are shown by crosses with error bars, models by asterisks. Right: measured spectra are shown by crosses and dotted lines, and model fits are indicated by solid lines (spectra). Errors on the spectra are indicated by a single black bar.\label{20N20S_spectra}}
\end{figure}

We also show full spectral fits for the 20$^{\circ}$N and 20$^{\circ}$S cases in Figure~\ref{20N20S_spectra}. This reinforces the clear difference in not only the limb darkening relation but also the spectral shape between the two latitudes, with the 20$^{\circ}$N spectra clearly seen to be flatter in shape. The Dii model (non-scattering NH$_4$SH cloud, scattering haze) clearly fits both sets of spectra well.

We can also produce a reasonable fit to the data with non-scattering grey cloud/scattering grey haze model E, although the best fit is achieved with an NH$_3$ or NH$_4$SH cloud. The fact that a grey cloud can also reproduce the data means that we still have no strong evidence for a particular cloud composition. This is reinforced by the fact that the best-fit particle sizes for the NH$_3$ and NH$_4$SH clouds are the sizes for which the extinction coefficient is relatively flat across the VIMS wavelengths (Figure~\ref{ext}), indicating that no significant absorption features due to cloud are present within the spectra. 

\section{Discussion}
Generally, latitudinal trends in gas abundance agree with the findings of \citet{fletcher11}, with a clear peak in NH$_3$ at the equator and both PH$_3$ and AsH$_3$ elevated in the southern hemisphere relative to the north. The retrieved haze optical depth is also reduced in the northern hemisphere for models where the haze is present. No significant trends in cloud particle size or base pressure are found as a function of latitude.

\subsection{Cloud and haze}
We found that, of the five cloud models tested, the best fit overall was obtained for variants of model D, with an NH$_4$SH deep cloud and NH$_3$ 2-$\upmu$m tropospheric haze (chosen for the relative spectral invariance of its scattering properties), although model C also provides a reasonable fit. Hazy models C and D are strongly favoured over haze-free models A and B. These results, coupled with observations of a tropospheric haze layer using other instruments and also Cassini/VIMS measurements on the dayside, lead us to conclude that the models including tropospheric haze should be favoured. However, for completeness, we here discuss the effect of the haze on the retrieval.

In general, the latitudinal trends in the retrieved values for gases are independent of the cloud model. The PH$_3$, NH$_3$ and AsH$_3$ VMRs are all slightly higher in the southern hemisphere for the haze-free case, perhaps indicating that gas absorption is compensating for the reduced opacity when the haze is removed. The PH$_3$ scale height is also larger for the haze-free case, indicating that more PH$_3$ absorption is required higher in the atmosphere. The differences are negligible in the northern hemisphere, where the retrieved haze optical depth is small anyway. The cloud optical depth is higher in the southern hemisphere for the haze-free models, again, suggesting that the haze provides significant opacity. The cloud optical depth for the hazy models is very consistent with latitude, implying that the haze is responsible for the large variation in 5 $\upmu$m brightness with latitude that is seen in Figure~\ref{leigh_cubes}. 

The best fit cloud base pressure was similar for all cloud models tested, occurring between 1.5 and 2.7 bar. It is possible to place a good constraint on this value as these pressures occur within the wings of the weighting function at 5 $\upmu$m, so the measurement is highly sensitive to the location of the cloud. The best-fit particle size is  larger than 1 $\upmu$m in all cases; the relative flatness of the extinction cross section for particles of 1, 3 and 10 $\upmu$m for both NH$_3$ and NH$_4$SH means it difficult to discriminate between these particle sizes, as all can produce a reasonable fit in most cases. We can exclude sub-micron-sized particles with high confidence. 

The best-fitting particle size for NH$_3$ clouds is consistently found to be 10 $\upmu$m, which is at the upper limit of the sizes tested,  but for NH$_4$SH the best-fit size varies as a function of latitude. However, the reduced $\chi^2$ for 1, 3 and 10 $\upmu$m is similar for all latitudes. A best-fit particle size of 1 $\upmu$m is associated with lower cloud optical depths, cloud base pressures and PH$_3$ volume mixing ratios. 1 $\upmu$m particles have stronger extinction at shorter wavelengths than 3$\upmu$m and 10$\upmu$m-sized particles (Figure~\ref{ext}), and this effect trades off with raising the cloud deck higher in the atmosphere. Whilst there is little evidence to favour any specific particle size above 1 $\upmu$m over any other, it is clear that particles with a flat extinction cross-section are favoured. This is is keeping with the fact that a grey cloud model also provides a reasonable fit to the spectra.

There is strong degeneracy between the PH$_3$ VMR and the cloud base pressure, with lower base pressures associated with lower PH$_3$ abundances and smaller scale heights. This effect is a cautionary reminder of the degeneracies present in problems like this, and the dependence of other retrieved values on details of the cloud model. 

The retrieved haze optical depth for models C, D and E is higher in the southern hemisphere compared with the northern hemisphere, which is consistent with the findings of \citet{fletcher11} that the haze optical depth increases in the southern hemisphere. It should be noted that there is significant degeneracy in the retrieval between the cloud and haze optical depths, with these parameters inversely correlated - so a higher retrieved haze optical depth can be offset to some extent by a lower cloud optical depth. This does not greatly affect retrievals of other parameters, but should be borne in mind when intepreting these results. 

In addition, the optical depth of the tropospheric cloud is highly dependent on the model used, and is generally slightly higher than that found by \citet{fletcher11} (1---2 instead of 0.1---2), although it is within the same range. The huge spread of values retrieved by \citet{fletcher11} for different cloud models indicates the dependency on the precise cloud model, and it is to this that we attribute the discrepancies. \citet{fletcher11} observe a small maximum in tropospheric cloud optical depth at around 20$^{\circ}$N. For the NH$_3$ cloud case (C), we see a slight decrease in optical depth towards the highest northern latitudes. The retrieved optical depth for the NH$_4$SH cloud is more variable, but this is due to degeneracies with the particle size. The optical depth for the grey cloud case (E) is also relatively uniform with latitude, so the main driver of the increased brightness in the northern hemisphere appears to be the tropospheric haze rather than the cloud. 
 
\subsection{Cloud composition}
The retrieved cloud base pressure over all latitudes and for all models is found to be between 1.5 and 2.7 bar, which is consistent with previous results (e.g. \citealt{fletcher11}, 1.8---3.0 bar; \citealt{roman13}, 1.75 bar) and lies between the predicted base pressures for NH$_3$ and NH$_4$SH clouds \citep{atreya05b}. Therefore, this result does not provide any evidence for us to favour one of cloud models C and D over the other, and may instead imply that the tropospheric cloud is formed from a composite of NH$_3$ and NH$_4$SH. Another possible interpretation is that the tropospheric haze corresponds to the predicted NH$_3$ cloud and the tropospheric cloud to the predicted NH$_4$SH cloud, with the formation pressures being slightly lower for both than those suggested in the literature. However, these data do not allow discrimination between the scenarios presented here, and it is difficult to see how this question can be resolved in the absence of in-situ measurements. 

Is it certain, however, that whatever the bulk composition of the tropospheric cloud it is not a pure species, as either pure NH$_3$ or pure NH$_4$SH of the sizes that provide the best fit would scatter a substantial amount of light. However, the models that provide the best overall fit are models for which the tropospheric cloud is forced to be non-scattering. If NH$_3$ or NH$_4$SH are present these species must be contaminated with something that darkens the individual particles and makes them more absorbing. Usually, dark contaminants of this kind might be expected to be photochemically produced, but this seems unlikely to be the case here as the haze is uncontaminated. Impurities may possibly be formed slightly below the haze, then drift downwards before coating the cloud particles. However, it is difficult to further elaborate on this scenario with the current lack of ground truth for Saturn.

\subsection{PH$_3$}
\label{ph3}
Our results agree with those of \citet{fletcher11} in finding that a knee pressure between 1.1 and 1.5 bar produces a better fit to the nearest-nadir spectral shape than that derived from the CIRS results, so results from the VIMS instrument are consistent with each other but not with measurements made at longer wavelengths. As discussed by \citet{fletcher11}, this discrepancy may be due to unresolved degeneracies in the retrievals for one of the instruments, which seems likely as there is clearly degeneracy between the retrieved phosphine abundance and the cloud model used in this work (see the difference made by the inclusion of haze, and the variation in particle size for NH$_4$SH).

If the PH$_3$ knee pressure really is around 1.3 bar instead of the 0.55 bar derived from CIRS, there must be a mechanism for depleting PH$_3$ above the 1.3 bar level. Photolysis is the obvious process, but photolysis of PH$_3$ is unlikely to occur this deep in Saturn's atmosphere (\citealt{fletcher09} and references therein). Turbulent mixing with PH$_3$-poor atmosphere higher up could also produce the effect.

We generally retrieve a somewhat higher PH$_3$ VMR than \citet{fletcher11} using cloud models C, D and E, which difference can again be attributed to differences in the details of the cloud model used,  but we do see a hint of the decrease in abundance going from the southern to the northern hemisphere. However, we don't see the peak at +10$^{\circ}$ that is hinted at in \citet{fletcher11}, most likely as a result of the much broader latitude regions we use. We retrieve a similar PH$_3$ fractional scale height to \citet{fletcher11}, also decreasing from the southern to the northern hemisphere. The higher deep abundances retrieved are more consistent with those derived from CIRS  observations \citep{fletcher09} than the results of \citet{fletcher11}.

\subsection{NH$_3$ and AsH$_3$}
\label{nh3_ash3}
The variation in retrieved NH$_3$ abundance as a function of latitude is consistent with the findings of \citet{fletcher11}, with an obvious peak at the equator. However, we retrieve deep abundances ($>1.0$ bar) that are typically a factor of two higher than those of \citet{fletcher11}.  Results are particularly discrepant in the equatorial peak, with an especially high abundance (a factor of 3 greater than found by \citealt{fletcher11} using a grey, non-scattering cloud model) retrieved using model E. 

The observed discrepancy in deep NH$_3$ abundance between these results and those of \citet{fletcher11} can most likely be attributed to the high degeneracy between the chosen cloud model, retrieved cloud properties and other model parameters. NH$_3$ is particularly affected as it absorbs over most of this spectral region, in much the same way as the cloud does. In addition, we average spectra over much broader bins in this work, which may also be a contributing factor. Measurements such as these, obtained over a relatively narrow range of wavelengths, are often subject to this kind of problem. In a future paper we aim to use dayside reflection spectra to inform our models of the tropospheric haze, which in conjunction with the results from this paper will further specify the cloud properties and thus should enable us to better constrain gas abundances in the deep atmosphere.

The AsH$_3$ abundances are very similar between the two analyses, except that we do not see the reduction in AsH$_3$ towards higher southern latitudes that is observed by \citet{fletcher11}. This may be due to the coarser binning making it impossible to resolve the decrease, or the complex cloud model degeneracies already mentioned. However, the retrieved abundances are the same within the error bars. 

\begin{table}
\centering
\begin{tabular}[c]{|c|c|c|}
\hline
Observation&Date&Integration Time (s)\\
\hline
CM1551785063 & 2007-03-5 & 120 \\
CM1551785788 & 2007-03-5 & 120 \\
CM1551786483 & 2007-03-5 & 120 \\
CM1551787152 & 2007-03-5 & 80 \\
CM1551787500 & 2007-03-5 & 80 \\
CM1551787847 & 2007-03-5 & 80 \\
CM1551788194 & 2007-03-5 & 80 \\
CM1551788541 & 2007-03-5 & 80 \\
CM1551788889 & 2007-03-5 & 80 \\
CM1551789236 & 2007-03-5 & 80 \\
CM1551789583 & 2007-03-5 & 80 \\
CM1551789931 & 2007-03-5 & 80 \\
CM1551791020 & 2007-03-5 & 120 \\
CM1551791680 & 2007-03-5 & 120 \\
CM1551792345 & 2007-03-5 & 120 \\
CM1551793030 & 2007-03-5 & 120 \\
CM1560840624 & 2007-06-18 & 320\\
CM1560842057 & 2007-06-18 & 160\\
CM1561470278 & 2007-06-25 & 160\\
CM1561470996 & 2007-06-25 & 160\\
CM1561471874 & 2007-06-25 & 160\\
CM1561472592 & 2007-06-25 & 160\\
CM1561473460 & 2007-06-25 & 160\\
CM1562652928 & 2007-07-9 & 320\\
CM1562654361 & 2007-07-9 & 160\\
\hline
\end{tabular}
\caption{List of 2007 data cubes used in the current research. \label{07datacubes}}
\end{table}

\subsection{Temporal trends}
\label{temporal}
\begin{figure}
\centering
\includegraphics[width=0.7\textwidth]{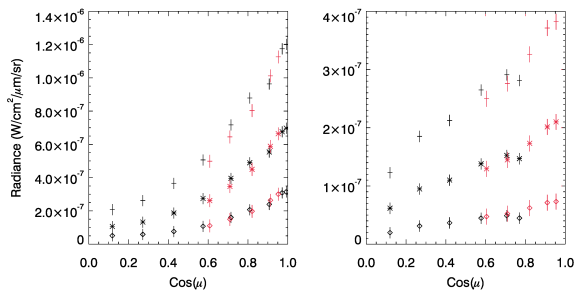}
\caption{Limb darkening curves for 2007 (black) compared with 2006 (red) for $\pm$20$^{\circ}$. The curves are clearly very similar between the two years and the latitudinal differences are preserved. The wavelengths are 5.12, 5.07 and 4.8 $\upmu$m moving down the plot. The radiances and limb darkening behaviour are very consistent between the two years.\label{20S07_data}}
\end{figure}

\begin{figure*}
\centering
\includegraphics[width=0.85\textwidth]{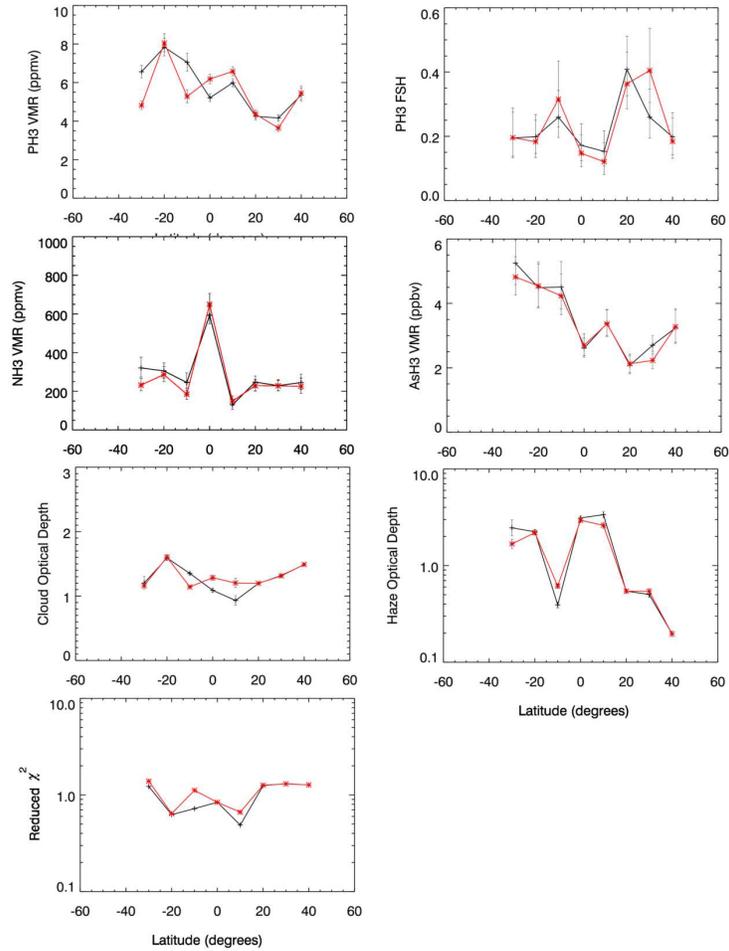}
\caption{Best-fit retrieved values for each latitude circle from the 2007 data for models Cii - black crosses and Dii - red stars, as Figure~\ref{resultsall}. Due to the differing geometry the emission angle range for -40$^{\circ}$ and 50$^{\circ}$ is reduced from the 2006 case, so we do not perform retrievals for these latitudes. However, the range for 0, 10, 20 and 30$^{\circ}$ is significantly increased. Hence, this is a good test of the validity of the model as it can clearly be applied to the 2007 data without modification.  \label{results_2007}}
\end{figure*}

We find that the striking hemispheric difference in the shape of the limb darkening curves is preserved into the following year, 2007. We examine 2007 data cubes listed in Table~\ref{07datacubes}. For comparison, we show the limb darkening curves at $\pm$20$^{\circ}$ for both years (Figure~\ref{20S07_data}). In addition, we perform a full limb darkening retrieval analysis using the best fit Cii and Dii models from the 2006 analysis, and obtain almost identical latitudinal trends. This is a good test of the model, since in the southern hemisphere the emission angles are typically higher in 2007 than in 2006, and in the northern hemisphere and at the equator the emission angle range is much larger. Results are shown in Figure~\ref{results_2007}.

\begin{figure}
\centering
\includegraphics[width=0.4\textwidth]{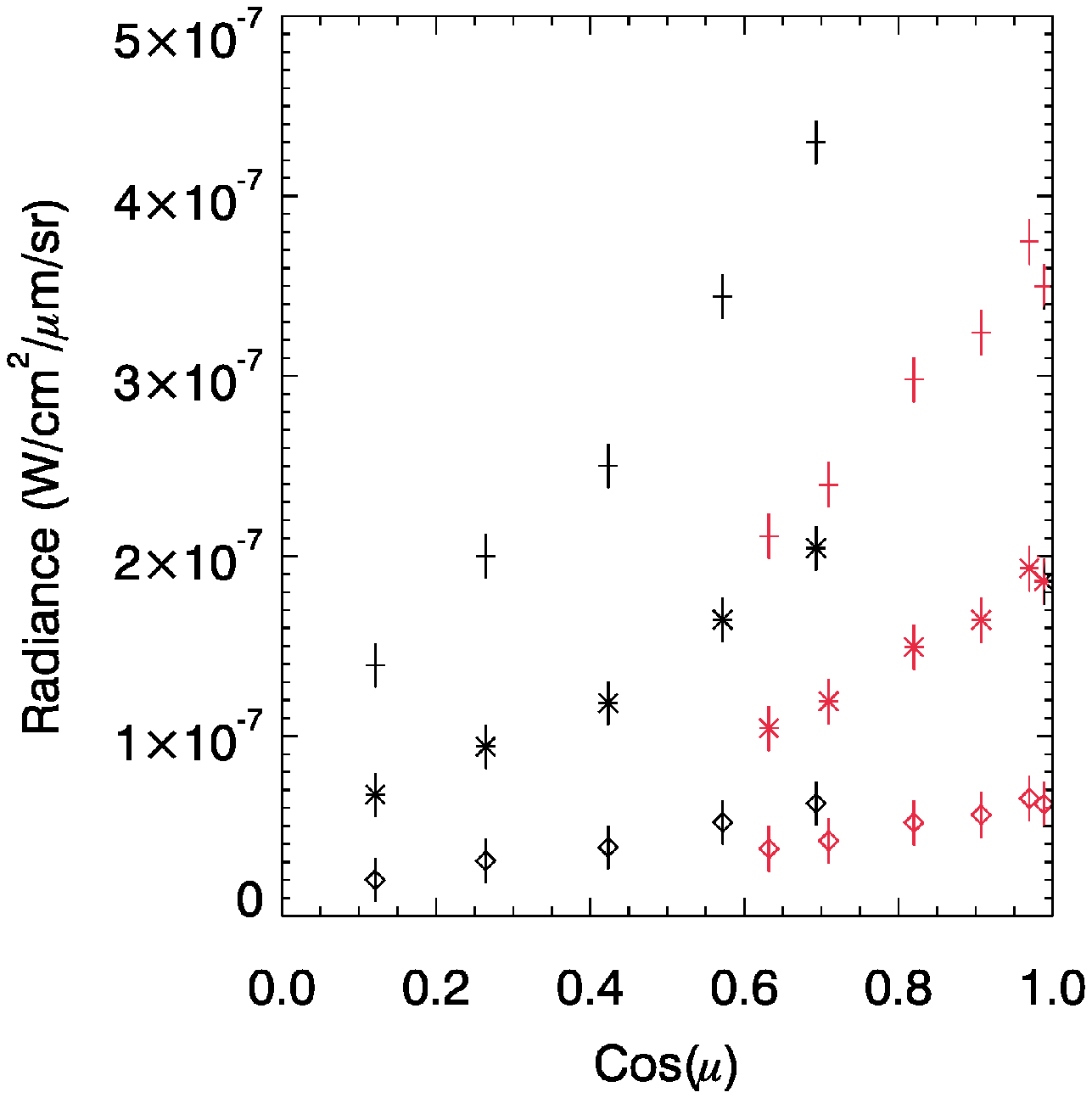}
\caption{Limb darkening curves for 2007 (black) compared with 2006 (red) for -10$^{\circ}$. At this latitude there is a clear difference between the two years, with the 2007 radiances being much brighter than 2006, despite the generally higher emission angles. This increased brightness is reflected in the retrieved results. \label{2007v2006_10S}}
\end{figure}

Some small differences in the retrieval results may be observed between the two years. The peak abundance of NH$_3$ at the equator is slightly reduced in 2007 from 2006. There is also a sharp decrease in retrieved haze optical depth at -10$^{\circ}$, which is not observed in 2006. Tests performed with different cloud and haze priors indicate that this is robust, despite the degeneracy between cloud and haze optical depths. An examination of the data for this latitude between the two years shows a strong brightening between 2006 and 2007, which is consistent with a clearing of the haze (Figure~\ref{2007v2006_10S}).

As stated above, unfortunately datasets with this wide emission angle range were not obtained with VIMS in subsequent years up to the 2010 storm, which significantly disrupted the northern hemisphere. However, as the storm is now dying down, any such data obtained towards the end of mission could be extremely useful, to compare the cloud structure during the current season on Saturn with the 2006---2007 epoch. We might expect to see increasing haze opacity in the northern hemisphere, and decreasing opacity in the south.

\section{Conclusions}
Building upon the near-nadir geometry work of \citet{fletcher11}, investigating the limb darkening behaviour of Saturn's clouds using the Cassini/VIMS instrument has uncovered further global trends in the cloud properties. There are significant hemispheric differences in the shape of the limb darkening curves, with much steeper limb darkening in the northern hemisphere during late northern winter, indicating that the cloud and haze must be less scattering overall in the northern hemisphere. This behaviour can be reproduced at all latitudes by a model with non-scattering tropospheric cloud and scattering tropospheric haze; as the haze optical depth decreases from the southern to the northern hemisphere, the non-scattering cloud dominates, meaning that the limb darkening curves are steeper in the north.

There is strong evidence for the presence of tropospheric haze from other instruments (e.g. see the \citealt{roman13} results from the ISS instrument), which is borne out by the VIMS retrievals presented here. However, as found by \citet{fletcher11} the problem is still highly degenerate and it is difficult to determine which hazy model is the best overall representation of Saturn's tropospheric cloud and haze. This is partly because clouds introduce several parameters into the model, and also because we do not observe any absorption features in the spectrum that are directly attributable to the cloud. Through simply comparing which of our model classes provides a good fit at the greatest number of latitudes, we find that a tropospheric, non-scattering NH$_4$SH cloud with a haze layer above is marginally favoured over other cloud models; however, given the high degeneracy of the problem we cannot rule out other models entirely with this dataset alone, and whatever the tropospheric cloud is made of it must contain a contaminant that significantly reduces the single-scattering albedo of the particles from that of the pure species. 

 In a future paper, we hope to utilize the visible range of VIMS to study the stratospheric haze and place further constraints on the tropospheric haze, which will help to resolve some of the questions raised in this paper. However, in order to fully break these degeneracies, it will be necessary to send future spacecraft with either higher resolution spectrometers (to differentiate unambiguously between the effects of cloud and absorption due to gaseous species) and/or descent probes to directly sample the tropospheric environment. 

\section{Ackowledgements}
JKB and PGJI acknowledge funding from the Science and Technology Facilities Council for this work. LNF is funded by a Royal Society University Research Fellowship. RSG also acknowledges the support of the Royal Society. We thank the two anonymous reviewers for their helpful and constructive comments on the manuscript.

\bibliographystyle{elsarticle-harv}
\bibliography{bibliography_saturn}

\label{lastpage}
\end{document}